\documentclass[twocolumn,showpacs,preprintnumbers,amsmath,amssymb]{revtex4}

\usepackage{graphicx}
\usepackage{dcolumn}
\usepackage{bm}

\begin{document}

\title{Better Synchronizability Predicted by Crossed Double Cycle}
\author{Tao Zhou}
\author{Ming Zhao}
\author{Bing-Hong Wang}
\email{bhwang@ustc.edu.cn}
\affiliation{%
Department of Modern Physics and Nonlinear Science Center,
University of Science and Technology of China, Hefei Anhui,
230026, PR China
}%

\date{\today}

\begin{abstract}
In this brief report, we propose a network model named crossed
double cycles, which are completely symmetrical and can be
considered as the extensions of nearest-neighboring lattices. The
synchronizability, measured by eigenratio $R$, can be sharply
enhanced by adjusting the only parameter, crossed length $m$. The
eigenratio $R$ is shown very sensitive to the average distance
$L$, and the smaller average distance will lead to better
synchronizability. Furthermore, we find that, in a wide interval,
the eigenratio $R$ approximately obeys a power-law form as $R\sim
L^{1.5}$.
\end{abstract}


\maketitle

Synchronization is observed in a variety of natural, social,
physical and biological systems\cite{Strogatz2003}, and has found
applications in a variety of field including communications,
optics, neural networks and
geophysics\cite{Pecora1990,Cuomo1993,Winful1990,Otsuka,Hansel1992,Vieira1999}.
The large networks of coupled dynamical systems that exhibit
synchronized state are subjects of great interest. In the early
stage, the corresponding studies are restricted to either the
regular networks\cite{Reg1,Reg2}, or the random
ones\cite{ran1,ran2}. However, recent empirical studies have
demonstrated that many real-life networks can not be treated as
regular or random networks. The most important two of their common
statistic characteristics are called small-world effect\cite{SWN}
and scale-free property\cite{SFN}. Therefore, very recently, most
of the studies about network synchronization focus on complex
networks, and find that the networks of small-world effect and
scale-free property may be easier to synchronize than regular
lattices\cite{syn3,syn5,Kwon2002}.

\begin{figure}
\scalebox{0.65}[0.65]{\includegraphics{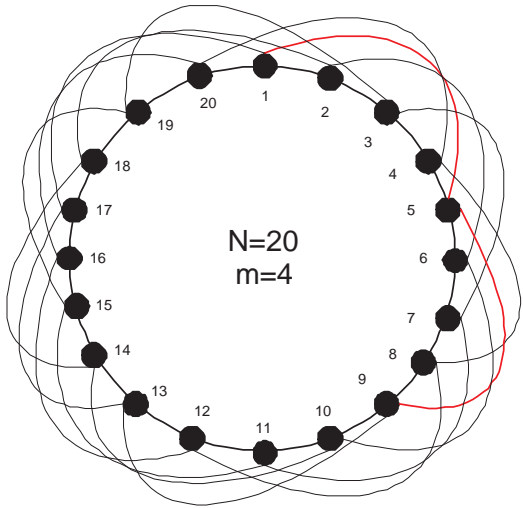}} \caption{(Color
online)The sketch maps of $G(20,4)$.}
\end{figure}

One of the ultimate goals in studying network synchronization is
to understand how the network topology affects the
synchronizability. In the simplest case (see below), the network
synchronizability can be well measured by the eigenratio $R$
\cite{master1,master2,Barahona2002,Pecora2005}, thus the above
question degenerates to understanding the relationship between
network structure and its eigenvalues. Since there are countless
topological characters for networks, a natural question is
addressed: what is the most important factor by which the
synchroizability of the system is mainly determined? Some previous
works indicated the average distance $L$\cite{ex1} is one of the
key factors. However, the consistent conclusion have not been
achieved\cite{Barahona2002,Gade2000,Lind2004,Nishikawa2003,Hasegawa2004}.
Another extensively studied one is the network heterogeneity,
which can be measured by the variance of degree distribution or
betweenness distribution\cite{betweenness1,betweenness2}. Some
detailed comparisons among various networks have been done,
indicating the network synchronizability will be better with
smaller heterogeneity\cite{Nishikawa2003,Hong2004,Zhao2005}.
However, a well-known counterexample is the regular networks with
homogeneous structure while displaying very poor
synchronizability. Because of the networks used for comparing in
previous studies are of both varying average distances and degree
variances, the strict and clear conclusions can not be achieved.
In addition, some researchers deem that the more intrinsic
ingredient leading to better synchronizability is the
randomicity\cite{Qi2003}, that is to say, the intrinsic reason
making small-world and scale-free networks having better
synchronizability than regular ones is their random structures.
Therefore, if one wants to clearly show how $L$ affects the
network synchronizability, he should investigate the networks of
different $L$ but the same degree variance. And if he wants to
assert that it is not the randomicity but smaller (or longer) $L$
resulting in the better synchronizability, the deterministic
networks are required.

In this brief report, we proposed a deterministic network model
named {\bf Crossed Double Cycles} (CDCs for short). The CDCs are
of degree variance equal to zero, and by adjusting the only
parameter $m$, named crossed length, the average distance of CDCs
can be changed. By using this ideal model, we demonstrate that the
smaller $L$ will result in better synchronizability, and provide a
useful method to enhance the synchronizability of nearest-neighbor
coupling networks.

\begin{figure}
\scalebox{0.82}[0.85]{\includegraphics{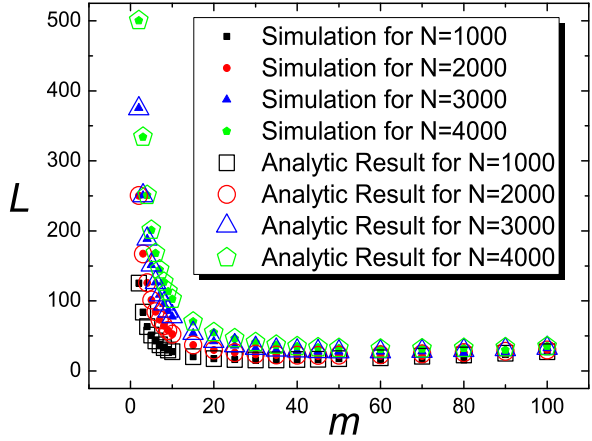}} \caption{(Color
online) The average distance of the CDCs. The black squares, red
circles, blue triangles and green pentagons represent the case of
$N=$1000, 2000, 3000 and 4000, respectively. The smaller solid
symbols and larger hollow symbols represent the simulation and
analytic results, respectively.}
\end{figure}

In network language\cite{Bondy1976}, the cycle $C_N$ denotes a
network consisting of $N$ vertices ${x_1,x_2,\cdots,x_N}$. These
$N$ vertices is arranged as a ring, and the nearest two vertices
are connected. Hence, $C_N$ has $N$ edges connecting the vertices
$x_1x_2,x_2x_3,\cdots,x_{N-1}x_N$, and $x_Nx_1$. The CDCs, denoted
by $G(N,m)$, can be constructed by adding two edges, called
crossed edges, to each vertex in $C_N$. The two vertices
connecting by a crossed edge are of distance $m$ in $C_N$. For
example, the network $G(N,3)$ can be constructed from $C_N$ by
connecting $x_1x_4,x_2x_5,\cdots,x_{N-1}x_2$, and $x_Nx_3$. And
the network $G(N,2)$ is isomorphic\cite{ex2} to a one-dimensional
lattice with periodic boundary conditions wherein each vertex
connects to its nearest and next-nearest neighbors. A sketch map
of $G(20,4)$ is shown in Fig. 1.

\begin{figure}
\scalebox{0.8}[0.85]{\includegraphics{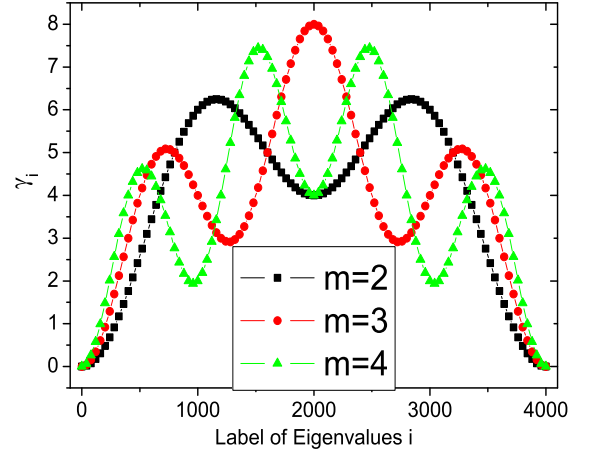}} \caption{(Color
online) Eigenvalues. The network size $N=4000$ if fixed, and the
black squares, red circles and green triangles denote the
numerical results of eigenvalues for the cases $m=2$, $m=3$, and
$m=4$, respectively. The corresponding curves represent the
analytical solutions.}
\end{figure}

\begin{figure}
\scalebox{0.8}[0.8]{\includegraphics{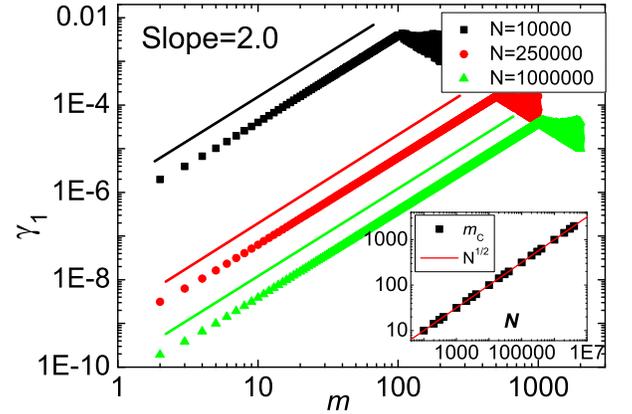}} \caption{(Color
online) The smallest nonzero eigenvalue $\gamma_1(\theta_1)$ vs
$m$. The black squares, red circles and green triangles denote the
cases $N=10000$, $N=250000$, and $N=1000000$, respectively. The
solid lines are of slope 2 for comparison. The inset shows the
cut-off point $m_c$ as a function of network size.}
\end{figure}

\begin{figure}
\scalebox{0.8}[0.8]{\includegraphics{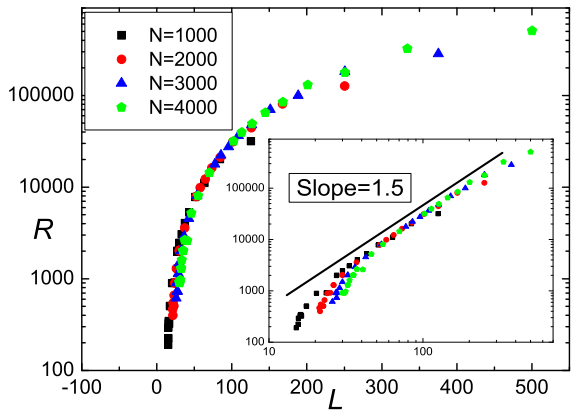}} \caption{(Color
online) $R$ vs $L$. The black squares, red circles, blue triangles
and green pentagons represent the case of $N=$1000, 2000, 3000 and
4000, respectively. The inset shows the same data in log-log plot,
indicating that the eigenratio $R$ approximately obeys a power-law
form as $R\sim L^{1.5}$. The solid line is of slope 1.5 for
comparison.}
\end{figure}

Clearly, all the vertices in $G(N,m)$ are of degree 4, thus the
degree variance is equal to 0. Furthermore, $G(N,m)$ is
vertex-transitivity, that is to say, for any two vertices $x$ and
$y$ in $G(N,m)$, there exists an automorphism mapping $\theta :
V(G)\rightarrow V(G)$ such that $y=\theta(x)$. The
vertex-transitivity networks are completely symmetric, which are
of particular practicability in the design of topological
structures of data memory allocation and multiple processor
systems\cite{Xu2001}.

Denote $L(k)$ the average distance of $C_{k+1}$, we have
\begin{equation}
    L(k)=\left\{
    \begin{array}{cc}
    \frac{k+2}{4} ,   &\mbox{$k$ is even}\\
    \frac{(k+1)^2}{4k} ,   &\mbox{$k$ is odd}\\
    \end{array}
    \right.
\end{equation}
For $N\gg m$, we assume $N$ can be exactly divided by $m$ and
denote $k=\frac{N}{m}$. Since $G(N,m)$ is vertex-transitivity, the
average distance of $G(N,m)$ is equal to the average distance
between vertex $x_1$ to all other vertices. The network $G(N,m)$
contains $k$ end-to-end $C_{m+1}$ as ${x_1x_2\cdots
x_{m+1},x_{m+1}x_{m+2}\cdots x_{2m+1},\cdots,x_{N+1-m}\cdots
x_Nx_1}$. Then, go from the vertex $x_1$ to a certain vertex $x_i$
can be divided to two processes. Firstly, travel through the
crossed edges to the nearest vertex that belongs to $x_i$'s cycle
mentioned above. Secondly, pass by a shortest path restricted in
this cycle to $x_i$. For example, the path from $x_1$ to $x_{10}$
in $G(20,4)$ is $x_1\rightarrow x_5 \rightarrow x_9 \rightarrow
x_{10}$. The first two edges are the crossed edges and identified
by red lines in Fig. 1, and the last edge is in the cycle
$x_9x_{10}x_{11}x_{12}x_{13}$. Hence one can obtain the average
distance of $G(N,m)$ for $N\gg m$ as:
\begin{equation}
L_G(N,m)=L(m)+L(k)-1.
\end{equation}
Figure 2 shows the simulation results of $L_G(N,m)$ for $N=1000,
2000, 3000, 4000$ and $m\leq 100$, which accurately agree with the
analytic ones.

In succession, we investigate the changes of CDCs'
synchronizability with $m$. Consider $N$ identical dynamical
systems (oscillators) with the same output function, which are
located on the vertices of a network and coupled linearly and
symmetrically with neighbors connected by edges of the network.
The coupling fashion ensures the synchronization manifold an
invariant manifold, and the dynamics can be locally linearized
near the synchronous state. The state of the $i$th oscillator is
described by $\textbf{x}^i$, and the set of equations of motion
governing the dynamics of the $N$ coupled oscillators is
\begin{equation}
\dot{\textbf{x}}^i=\textbf{F}(\textbf{x}^i)+\sigma\sum_{j=1}^NG_{ij}\textbf{H}(\textbf{x}^j),
\end{equation}
where $\dot{\textbf{x}}^i=\textbf{F}(\textbf{x}^i)$ governs the
dynamics of individual oscillator, $\textbf{H}(\textbf{x}^j)$ is
the output function and $\sigma$ is the coupling strength. The
$N\times N$ Laplacian $\textbf{G}$ is given by
\begin{equation}
    G_{ij}=\left\{
    \begin{array}{cc}
    k_i   &\mbox{for $i=j$}\\
     -1    &\mbox{for $j\in\Lambda_i$}   .\\
     0    &\mbox{otherwise}
    \end{array}
    \right.
\end{equation}

Because of the positive semidefinite of $\textbf{G}$, all the
eigenvalues of it are nonnegative reals and the smallest
eigenvalue $\theta_0$ is always zero, for the rows of $\textbf{G}$
have zero sum. Thus, the eigenvalues can be ranked as
$\theta_0\leq\theta_1\leq\cdots\leq\theta_{N-1}$. The ratio of the
maximum eigenvalue $\theta_{N-1}$ to the smallest nonzero one
$\theta_1$ is widely used to measure the synchronizability of the
network\cite{Barahona2002,master1}, if the eigenratio
$R=\theta_{N-1}/\theta_1$ satisfies
\begin{equation}
R<\alpha_2/\alpha_1,
\end{equation}
we say the network is synchronizable. The right-hand side of this
inequality depends only on the dynamics of individual oscillator
and the output function\cite{Barahona2002}, while the eigenratio
$R$ depends only on the Laplacian $\textbf{G}$. $R$ indicates the
synchronizability of the network, the smaller it is the better
synchronizability and vice versa. In this brief report, for
universality, we will not address a particular dynamical system,
but concentrate on how the network topology affects eigenratio
$R$.

Since the Laplacian for any CDC is shift invariant, the
eigenvalues can be calculated from a discrete Fourier transform of
a row of the Laplacian matrix \cite{Pecora2005}. Denote $\gamma_i$
$(i=0,1,\cdots,N-1)$ the $i$th eigenvalue\cite{ex3}, it reads
\begin{equation}
\gamma_i=2\left(2-\texttt{cos}\frac{2\pi
i}{N}-\texttt{cos}\frac{2\pi im}{N}\right).
\end{equation}
Figure 3 shows the numerical results of eigenvalues, which
accurately agree with the analytical solutions. Clearly, for odd
$m$ and even $N$, the maximal eigenvalue is
$\theta_{N-1}=\gamma_{N/2}=8$. However, if $m$ is even or $N$ is
odd, the maximal eigenvalue $\theta_{N-1}$ is smaller than 8 and
can not be expressed sententiously. The smallest nonzero
eigenvalue $\theta_1$ equals $\gamma_1$, and can be approximately
obtained under the condition $N \gg m$
\begin{equation}
\theta_1=\gamma_1\approx \frac{4\pi^2}{N^2}(1+m^2).
\end{equation}
Figure 4 reports $\gamma_1$ as a function of $m$, which scales as
$m^2$, as predicted by Eq. (7), before reaching a cut-off point
$m_c$. The numerical value of $m_c$ is also shown in the inset of
Fig. 4, which accurately obeys the form $m_c=\sqrt{N}$.

Because of the cut-off in $\theta_1$, the fluctuations in
$\theta_{N-1}$ for even $m$, and the relatively complex
relationship between $m$ and $L$, we can not obtain a
straightforward expression to comprehensively depict the
relationship between $R$ and $L$. In Fig. 5, we only report the
numerical results about how the average distance affects the
network synchronizability. One can see clearly, the network
synchronizability is very sensitive to the average distance; as
the increase of $L$, the eigenratio $R$ sharply spans more than
three magnitudes. And the network synchronizability is remarkably
enhanced by reducing $L$. When the crossed length $m$ is not very
small or very large (comparing with $N$), the networks with the
same average distance have approximately the same
synchronizability no matter what the network size is. More
interesting, the calculated results indicate that the eigenratio
$R$ approximately obeys a power-law form as $R\sim L^{1.5}$ in a
wide interval of $L$ (see the inset of Fig. 5).

To sum up, we propose an ideal network model, and investigate its
synchronizability. The results indicate that the average distance
is an important factor affecting the network synchronizability
greatly. The smaller average distance will lead to better
synchronizability. This is similar to the communication systems,
wherein the average distance is one of the most important
parameters to measure the transmission delay (or time delay)
encountered by a message travelling through the network from its
source to destination, and the smaller average distance means
higher efficiency for homogeneous networks. Very recently, by
numerical studies, some authors think that there may exist some
common features between dynamics on communication networks
(traffic and diffusion) and network
synchronization\cite{Zhao2005,Motter2005,Chavez2005,Yin2005}.
Since in the former dynamics shorter $L$ will lead to greater
throughput and fast spread, the underlying common features provide
a possible explanation why shorter average distance corresponding
to better synchronizability.

The CDCs are natural extensions of the lattice of nearest
neighbors, they are symmetric and with better synchronizability,
thus have great potential in the applications for designing of
topological structures of distributed processing systems, local
area networks, data memory allocation and data alignment in single
instruction multiple data processors\cite{Xu2001}. In fact, the
processor network of one kind of the earliest parallel processing
computers is $G(16,4)$\cite{Barnes1968}. Besides in
synchronization, the cross method also has been applied in
communication systems. For example, the crossed cubes have much
larger throughput than hypercubes thus are widely used in
designing parallel computing
networks\cite{Efe1991,Kulasinghe1995}.

Supported by the NNSFC under Grant No. 10472116, 70471033 and
70571074.


\begin{thebibliography} {Reviews1}
\bibitem{ex1} In a network, the distance between two vertices is
defined as the number of edges along the shortest path connecting
them. The average distance $L$ of the network, then, is defined as
the mean distance between two vertices, averaged over all pairs of
vertices.
\bibitem{ex2} A network $G$ is isomorphic to a network $H$ if
there exists a bijective mapping $\theta : V(G)\rightarrow V(H)$
satisfying the adjacency-preserving condition: $(xy)\in E(G)
\Longleftrightarrow (\theta(x)\theta(y))\in E(H)$, where
$V(\bullet)$ and $E(\bullet)$ denote the sets of vertices and
edges, respectively. Popularly speaking, two networks are
isomorphic means they have the same structure.
\bibitem{ex3} Note that, $\theta_0,\theta_1,\cdots,\theta_{N-1}$ is the sort ascending of $\gamma_0,\gamma_1,\cdots,\gamma_{N-1}$.
\end{thebibliography}
\end{document}